**Anomalous vortex shape in a frustrated superconductor hosting chiral multicomponent order parameters**

Yuita Fujisawa[1,2], Anjana Krishnadas[1], Tomonori Nakamura[1], Chia-Hsiu Hsu[1,3], Yen-Yu Lai[4], Barnaby R. M. Smith[1], Markel Pardo-Almanza[1], Yukiko Obata[1,5], Dyon van Dinter[1], Guoqing Chang[4], Chun-Liang Lin[4], Robert. Joynt[6,7], Yuki Nagai[8,9], Tadashi Machida[10], Yoshinori Okada[1,11]

*[1]Quantum Materials Science Unit, Okinawa Institute of Science and Technology (OIST), Okinawa 904-0495, Japan.*

*[2]Hiroshima Synchrotron Radiation Centre, Hiroshima University, Higashi-Hiroshima, Hiroshima 739-0046, Japan.*

*[3]Division of Physics and Applied Physics, School of Physical and Mathematical Sciences, Nanyang Technological University, 637371 Singapore.*

*[4] Department of Electrophysics, National Yang Ming Chiao Tung University, Hsinchu 300, Taiwan.*

*[5]Department of Physics, Kanazawa University, Ishikawa 920-1192, Japan.*

*[6]Department of Physics, University of Wisconsin-Madison, 1150 University Avenue, Madison, WI 53706, USA.*

*[7]Theoretical Sciences Visiting Program, Okinawa Institute of Science and Technology Graduate University, Onna, 904-0495, Japan*

*[8]Information Technology Center, The University of Tokyo, Chiba 277-0882, Japan.*

*[9]CCSE, Japan Atomic Energy Agency, 178-4-4, Wakashiba, Kashiwa, Chiba, 277-0871, Japan.*

*[10]RIKEN Center for Emergent Matter Science, Saitama, 351-0198, Japan.*

*[11]JST, PRESTO, 4-1-8 Honcho, Kawaguchi, Saitama, 332-0012, Japan.*

Multicomponent chiral superconductors can host spatially separated superconducting components within a single Abrikosov vortex, leading to unconventional, anisotropic vortex shapes forbidden in conventional single-component superconductors. Here, using spectroscopic scanning tunneling microscopy at 0.3 K, we investigate the mixed state of the spinel superconductor $LiTi_2O_4$ ($T_c \approx 13$ K). In the low-field regime, where vortex–vortex interactions are minimized, and the intrinsic single-vortex shape is preserved, reasonably isolated triangular Abrikosov vortices are directly visualized with high statistical significance. Remarkably, the vortex orientation is locked to crystallographic domains rather than the magnetic-field direction, revealing a hidden chirality selectivity embedded in the zero-field electronic state.

Understanding hidden electronic ordering and its symmetry in superconductors—spanning both long- and short-range correlations—is a central challenge in quantum materials science [1,2,3,4,5,6,7,8,9,10,11,12,13,14,15,16,17,18,19,20,21]. A particularly powerful route toward this goal is the visualization of Abrikosov vortices (AVs), whose shapes, encoded in the spatial distribution of low-energy quasiparticle states, can directly reflect the internal symmetry of unconventional superconductivity. To date, however, most experimental studies have focused on superconductors described by single-component order parameters [22,23,24,25,26]. In contrast, a striking theoretical prediction that remains experimentally unexplored is that superconductors hosting a chiral multicomponent order parameter (CMOP), $\Psi\pm = \Psi_1 \pm i\Psi_2$, can support unconventional vortex textures beyond the conventional paradigm [27,28,29].

$LiTi_2O_4$ (LTO), with a formal $3d^{0.5}$ electronic configuration on the Ti pyrochlore lattice, is a unique transition-metal spinel superconductor with a relatively high critical temperature (Tc ≈ 13 K) [**Fig. 1(a)**] [30,31,32,33,34,35,36]. Its density-functional-theory-based electronic structure is shown in **Fig. 1(b)**. While the closely related $MgTi_2O_4$ (MTO) is a $3d^1$ Mott insulator exhibiting long-range spin, orbital, and lattice order accompanied by long-range chiral structures along the (100) direction [37]. LTO is different in that highly frustrated and highly degenerate electronic states are suggested to survive in the anomalous metallic regime above Tc [**Fig. 1(c)**] [34,38]. Importantly, it has recently been shown that (111)- and (−1−1−1)-oriented LTO films provide atomically flat surfaces suitable for spectroscopic-imaging STM studies [**Fig. 1(d)**]. These orientations have been attracted broad attention in condensed matter systems [39]. In describing the electronic structure of such films, it is natural to start from singly degenerate out-of-plane and doubly degenerate in-plane eigenstates, rather than the triply degenerate $t_{2g}$ manifold ($d_{xy}$, $d_{yz}$, $d_{zx}$), even though the bulk crystal remains cubic. Within this framework, the doubly degenerate in-plane states may be expressed in a chiral basis with opposite angular momenta, allowing for reflection of the local in-plane rotational symmetry. The symmetry of the superconducting state is expected to be intimately connected to the underlying crystal symmetry as reflected in the normal-state electronic structure [40], Thus, LTO provides an attractive platform for exploring chiral superconductivity emerging from geometrically frustrated 3d-electron systems.

Here, we perform *in situ* spectroscopic scanning tunneling microscopy on superconducting $LiTi_2O_4$ films, which host domain structures composed of coexisting regions with (111) and (−1−1−1) orientations normal to the film plane (see **Supplementary Note 1,2** for Methods and domain structure characterization, respectively). As shown in **Fig. 1(e)**, the superconducting gap measured at 0.3 K and zero magnetic field is nearly isotropic and nodeless across the domains. However, this does not necessarily indicate a conventional s-wave superconducting state. In a multicomponent chiral superconductor, the relative phase rotation between $\Psi_1$ and $\Psi_2$ can produce a nearly isotropic total gap amplitude, $|\Psi\pm| = \sqrt{(|\Psi_1|^2 + |\Psi_2|^2)}$, even when the individual components themselves contain nodes. By contrast, a nonchiral combination such as $\Psi_1 + \Psi_2$ generally enhances gap anisotropy because of partial cancellation between the two order parameters arising from their different signs. Under an applied magnetic field, however, the multicomponent chiral nature of the superconducting state can manifest itself through the emergence of an unconventional vortex structure. Instead of the conventional sixfold-symmetric vortices expected for single-component superconductivity based on the calculated electronic structure [**Fig. 1(f)**], we observe triangular Abrikosov vortices (tr-AVs) [**Fig. 1(g)**]. Moreover, the in-plane orientation of the tr-AVs rotates across crystalline domain boundaries [black line in **Fig. 1(g)**]. Based on these observations, we propose that LTO hosts an underlying chiral multicomponent superconducting order parameter, together with a hidden chirality selectivity already embedded in the zero-field electronic state.

Focusing on individual vortices, we introduce a method to quantify the vortex shape through the spatial structure of the zero-bias conductance. For LTO, the lower and upper critical fields are $B_{c1} \approx 30$ mT and $B_{c2} \approx 11$ T, respectively [41]. At B = 0.5 T, spatial maps of g(0)/g(4.5 mV) reveal a clear triangular vortex shape [**Fig. 2(a)**]. Spectra taken along a line through the vortex center show well-defined bound states [**Fig. 2(b,c)**], indicating that the system is essentially in the clean limit, enabling the investigation of the superconducting nature of LTO without significant impact from disorder. Using the normalized conductance map [**Fig. 2(a)**], we construct a polar intensity plot [**Fig. 2(d)**] and integrate the conductance up to 30 nm (approximately three times the coherence length, $\xi \approx 10$ nm [41]) to optimize the signal-to-noise ratio while minimizing contamination from neighboring vortices [**Fig. 2(e)**; **Supplementary Note 3**]. Fourier analysis on the obtained integrated polar conductance curve yields a robust threefold component, $I_{C3}$ [**Fig. 2(f)**].

To examine whether the tr-AVs can be understood within a conventional framework, we first consider the possible influence of surface-related effects. If the triangular vortex shape originated primarily from local surface structures, a clear correlation to the surface morphology would be expected. However, such behavior is not observed. Neither the positions nor the shapes of the vortices show any systematic correlation with surface features, including step edges and triangular islands within a single crystalline domain. We next consider whether anomalous surface electronic states could be responsible for the observed tr-AVs. Previous surface-sensitive ARPES measurements on $LiTi_2O_4$ films demonstrated that the surface electronic structure closely reflects the bulk band structure [34]. Consistently, the superconducting gap measured by surface-sensitive STM/STS agrees well with the bulk superconducting gap. Overall, these observations indicate that surface-related effects are unlikely to play a major role in the formation of the tr-AVs, consistent with the relatively long coherence length of ~10 nm, which is too large for local surface anomalies to dominantly determine the superconducting vortex shape.

As the next step, we examine whether the observations can be understood within a conventional single-component description of bulk superconductivity. In such a framework, vortex-core anisotropy is governed by $v_F(k)/|\Delta(k)|$, which remains invariant under $k \rightarrow -k$. Here, $v_F$ and $\Delta$ denote the Fermi velocity and superconducting pairing gap, respectively. While elliptical or hexagonal distortions are symmetry-allowed, a net threefold anisotropy cannot arise from a single complex order parameter preserving inversion symmetry. Accordingly, as shown in **Fig. 1(f)**, the expected vortex shape in LTO remains sixfold symmetric as long as a single-component order parameter is assumed.

Taken together, neither surface-driven mechanisms nor conventional single-component bulk descriptions can naturally account for the observed tr-AVs. The need for a distinct underlying mechanism is further supported by comparison with conventional superconductors, in which triangular-shaped vortices have not been observed despite the presence of analogous $C_3$ near-surface lattice symmetry and triangular crystalline islands, similar to those found on the LTO(111) surface [42,43].

The essential implication of the CMOP-based picture is that vortex shapes forbidden in any single-component description can emerge in multicomponent chiral superconductors. In this framework, the vortex structure is determined by a balance between two competing energy considerations. One tendency is to spatially separate the singular regions where $|\Psi_1(\mathbf{r})|$ and $|\Psi_2(\mathbf{r})|$ have zeros, which helps to minimize condensation-energy loss by maintaining a relatively large total superconducting amplitude $|\Psi(r)|$ [**Figs. 2(g–h)**]. The opposing tendency is to keep these singular regions locked in space, which also minimizes the gradient energy cost (**Supplementary note 4**). As a result of this competition, $|\Psi_1(\mathbf{r})|$ and $|\Psi_2(\mathbf{r})|$ form a multiple-core-like structure within a single vortex, giving rise to a tr-AV [**Figs. 2(g–h)**]. Importantly, in the present case, conceptually analogous competition emerges within a single vortex itself. Note that even if multiple core-like structures exist within a single tr-AV, the total magnetic flux carried by the vortex is still expected to correspond to a single flux quantum, $\phi_0$, as in a conventional Abrikosov vortex [27]. This phenomenology leading to tr-AV is fundamentally different from that of a conventional single-component superconductor, where a vortex is described by a single order parameter whose phase winds by $2\pi$ around the vortex core. In that case, the phase winding generally necessarily produces a single point at which the order-parameter amplitude vanishes, and the vortex shape is more isotropic though the crystalline anisotropy that follows $v_F(k)/|\Delta(k)|$ can still be present (see **Supplementary Note 4** for more details).

At this point, multiple pairing possibilities can be considered, including s-, p-, d-, or even f-wave channels [16,33,44,45]. However, within the present experimental resolution and considering the primary scope of this study, clarifying the detailed microscopic nature of the order parameters—including whether $\Psi_1$ and $\Psi_2$ originate from identical or distinct representations—is beyond the main focus of this work [46]. Nevertheless, our B=0 data [**Fig. 1(e)**] should be considered as strongly constraining any proposed model to have relatively isotropic pairing gap amplitudes.

As the next step, we discuss the nature of the domains and domain boundaries. Although STM images predominantly visualize the Ti triangular-lattice-terminated layer [32], two types of domains are recognized [**Figs. 3(a–cc)**]. This domain corresponds to Ti triangular-lattice-terminated surfaces viewed from opposite sides of the bulk crystal, meaning, from the surface point of view, domain A and domain B can be regarded as two regions oriented along the corresponding (111) and (-1-1-1) directions [**Fig. 3(d)**]. Note that although the stacking sequence along the out-of-plane direction inverts the Fermi surface at finite kz, whose direction is commonly set between the two domains [**Fig. 3(e)**], the $k_z$-integrated electronic states are expected to remain identical [**Fig. 3(f)**]. Nevertheless, this hidden selectivity does not produce a clear spectroscopic anomaly in quasiparticle tunneling at zero magnetic field. Indeed, nearly identical g(E)/g(4.5 mV) spectra and the absence of pronounced in-gap states are observed at 0.3 K both inside the domains and across the domain boundaries (DBs) [**Fig. 3(g)**]. At the phenomenological level, adjacent domains separated by a DB therefore remain electronically coherent as long as B = 0.

In striking contrast, under an applied magnetic field, the neighboring domains become clearly distinguishable. **Fig. 3h–j** present zero-bias conductance maps (**Fig. 3h**) together with histograms of $I_{C3}$ (**Fig. 3i**) and $\Theta_0$ (**Fig. 3j**) acquired under magnetic fields of 0.2, 0.5, and 1.0 T. These measurements further establish the robust observation of tr-AVs, confirming more than 80 vortices across the investigated field range. Furthermore, within a single domain, the orientations of the tr-AVs are uniformly locked, whereas their orientation rotates by 180° across the adjacent domain boundary (black lines in **Fig. 3h**).

Remarkably, the tr-AV orientation does not reverse when the magnetic-field polarity is inverted. **Figures 4(a–c)** show a topographic image [**Fig. 4(a)**] together with zero-bias conductance maps measured at $B$ = +0.5 T [**Fig. 4(b)**] and $B$ = −0.5 T [**Fig. 4(c)**]. For both field directions, the primary observation is the magnetic-field-orientation independence of the tr-AV, while only a slight polarization dependence is observed. This major feature immediately indicates that the twofold selectivity is linked to the nature of the crystalline domain. If the system distinguishes the out-of-plane polarity, the in-plane $C_3$ axis can be selectively fixed, allowing each domain to determine the orientation of the triangular Abrikosov vortices [**Fig. 3(e)**]. At present, however, the microscopic mechanism capable of pinning such a triangular direction within the nearly cubic crystal structure remains unclear [**Fig. 3(f)**]. Moreover, domain-dependent differences in the electronic structure, including those in the anomalous metallic state either with or without a magnetic field, have so far remained experimentally elusive. As emphasized above, surface-related effects are unlikely to play a dominant role in the formation of the tr-AVs. Nevertheless, our results suggest that a hidden chirality selectivity is already embedded in the zero-field electronic state.

The phenomenological picture that guides the collective interpretation of our observations is summarized schematically in **Fig. 4(d)**. In this picture, neighboring domains possess opposite hidden tendencies toward Ψ+(-) and Ψ−(+) already at B = 0, while the two chiral states remain electronically inactive in zero field, consistent with the absence of a clear pairing-gap anomaly across the DBs. Once the B-field is turned on, regardless of orientation, chirality becomes electronically active, locking the tr-AVs' in-plane orientation. While specifying the chirality Ψ+ and Ψ- in each domain is not possible at this stage, the domain boundary inverts the chirality between Ψ+ and Ψ−, an intriguing consequence of this picture that will be further developed elsewhere.

In summary, the observation of triangular Abrikosov vortices provides strong evidence for chiral multicomponent superconductivity in $LiTi_2O_4$. Also, a hidden chirality selectivity already embedded in the zero-field electronic state is poposed. Our findings open a broad avenue for exploring how analogous phenomena evolve as the electronic filling is tuned between the $3d^0$ and $3d^1$ limits, particularly in connection with anomalous metallic and superconducting states [**Fig. 1(c)**]. More broadly, our results establish epitaxial spinel-oxide thin films as a promising platform for investigating exotic superconductivity in strongly correlated 3d-electron systems on geometrically frustrated lattices and nanoscale structures, including heterointerfaces.

**Acknowledgment**

We thank the University of the Ryukyus Research Facility Center for the use of the helium liquefier.

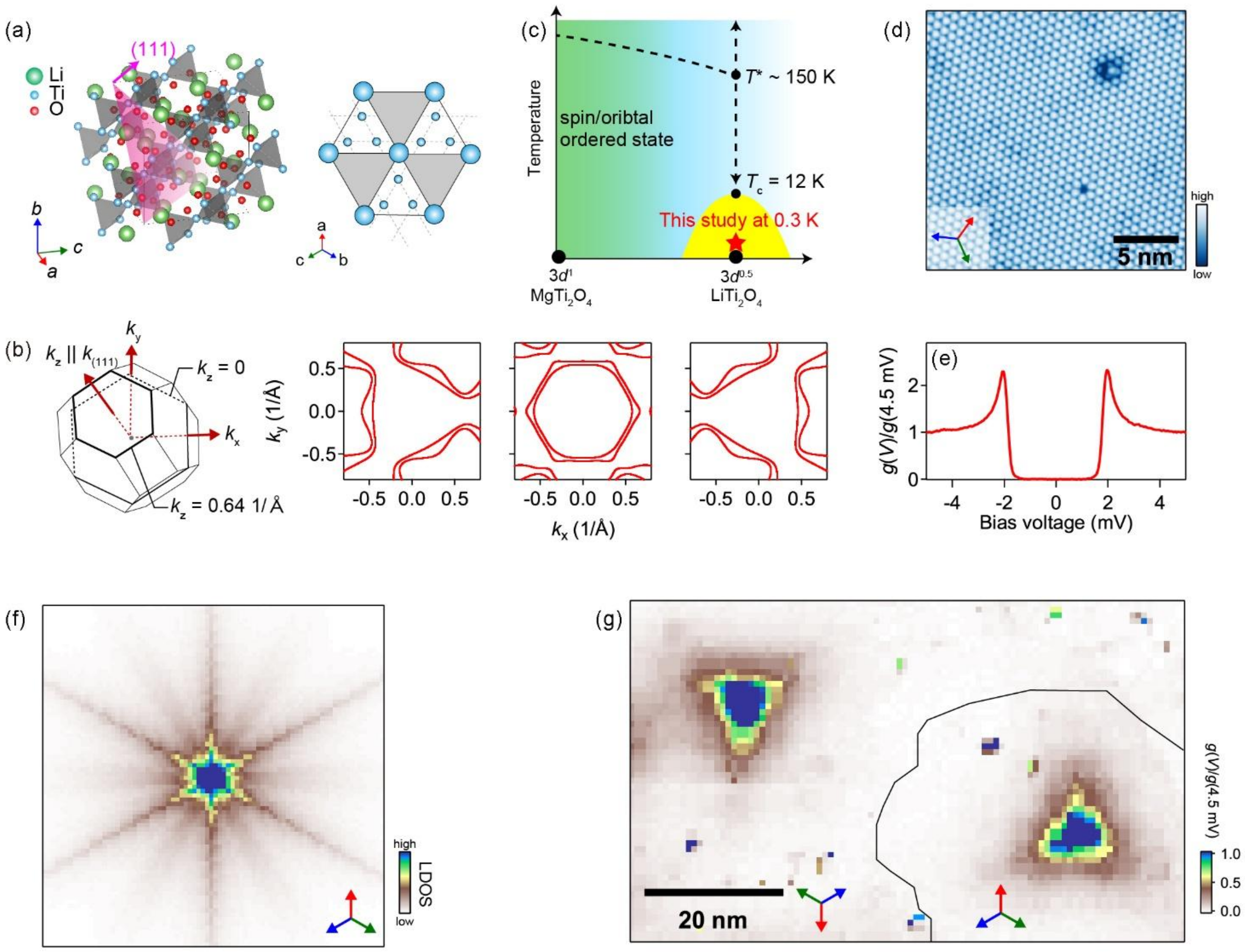

**Figure 1. Emergent triangular vortex state in the spinel superconductor $LiTi_2O_4$.**

(a) Crystal structure of LiTi2O4 (LTO) and schematic of the Ti pyrochlore sublattice.

(b) Brillouin zone of LTO and calculated Fermi surfaces at $k_z$ = -0.3, 0, 0.3 1/Å.

(c) Schematic phase diagram of the $MgTi_2O_4$–$LiTi_2O_4$ system.

(d) Atomically resolved STM topographic image of the $LiTi_2O_4$ surface. Setup: 800 mV, 100 pA.

(e) Tunneling conductance spectrum measured at 0.3 K. Setup: +7 mV, 1 nA; lock-in: 0.2 mV, 961 Hz.

(f) Simulated zero-bias conductance map acquired under magnetic field, revealing a vortex with pronounced sixfold anisotropy.

(g) Large-scale zero-bias conductance map showing spatially separated triangular vortex cores. The black line indicates the domain boundary. Setup: 5 mV, 500 pA; lock-in: 0.1 mV, 961 Hz.

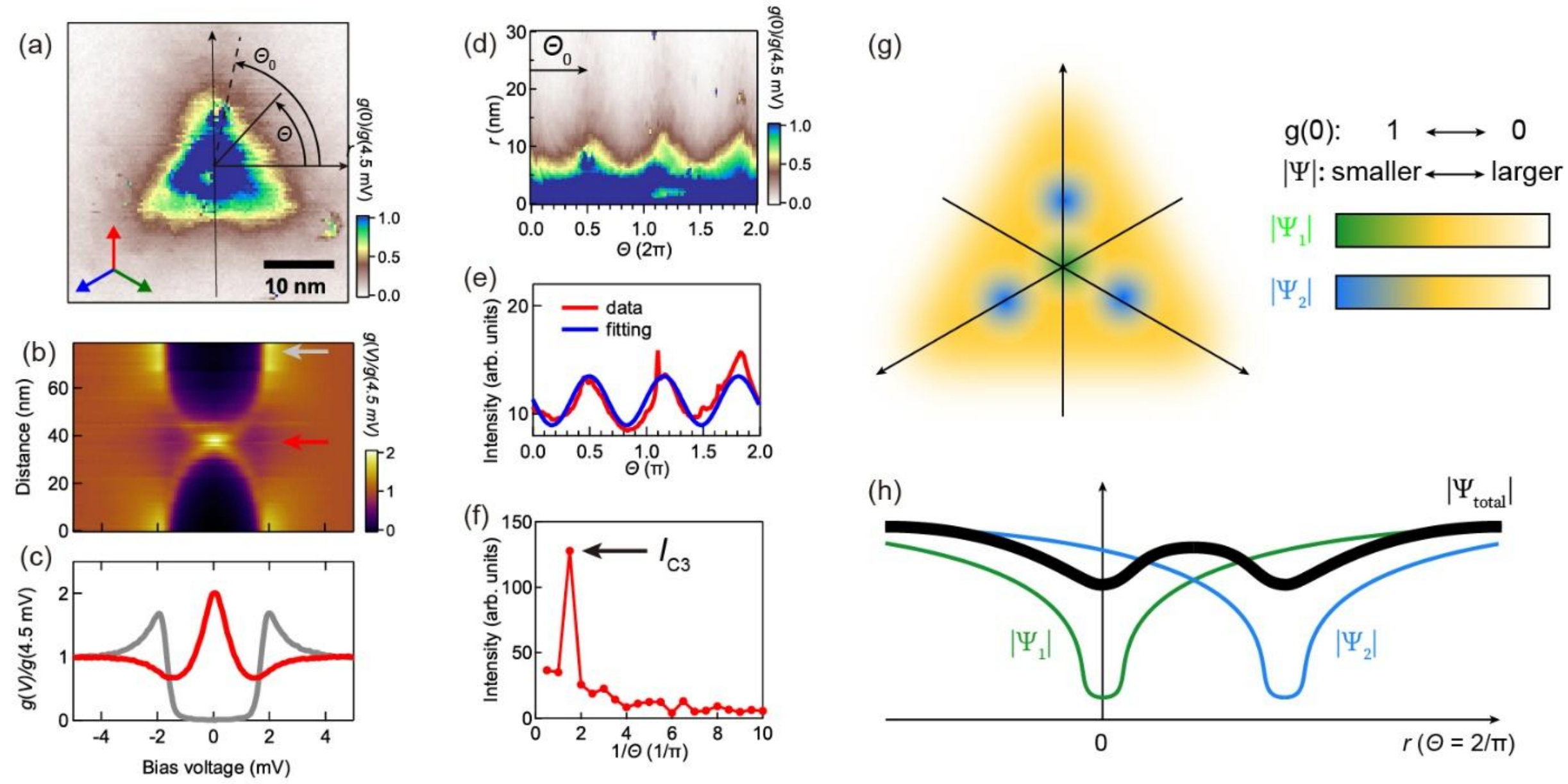

**FIG. 2. Triangular Abrikosov vortices in LTO.**

(a) Zero-bias conductance (ZBC) map showing a triangular Abrikosov vortex. Setup: 5 mV, 500 pA; lock-in: 0.1 mV, 961 Hz.

(b) Spectral evolution across the black line in (a). Red arrows mark the vortex-core states.

(c) g(V)/g(4.5 mV) spectra taken at the vortex core indicated by the red arrow in (b) (red) and away from the vortex (gray).

(d,e) Polar plot of the ZBC map (d) and integrated intensity from r = 0 to 30 nm (e), as a function of θ. Here, the polar angle θ is defined with respect to the horizontal axis and increases anticlockwise as in (a). The blue curve in (e) is a fitting curve using the $\cos(3\theta + \theta_0)$ factor, assuming three-fold symmetry. Here, $\theta_0$ corresponds to the vertex orientation angle for tr-Avs, and for technical convenience, $\theta_0$ is constrained to the ranges +60° to +120° and -60° to -120°. Note that (a) corresponds to the former case.

(f) Fourier transformed profile of (e).

(g) Schematic illustration of the spatially separated vortex-core structure composed of multiple superconducting components.

(h) Schematic spatial evolution of the superconducting order parameters and the total order parameter amplitude across the vortex core.

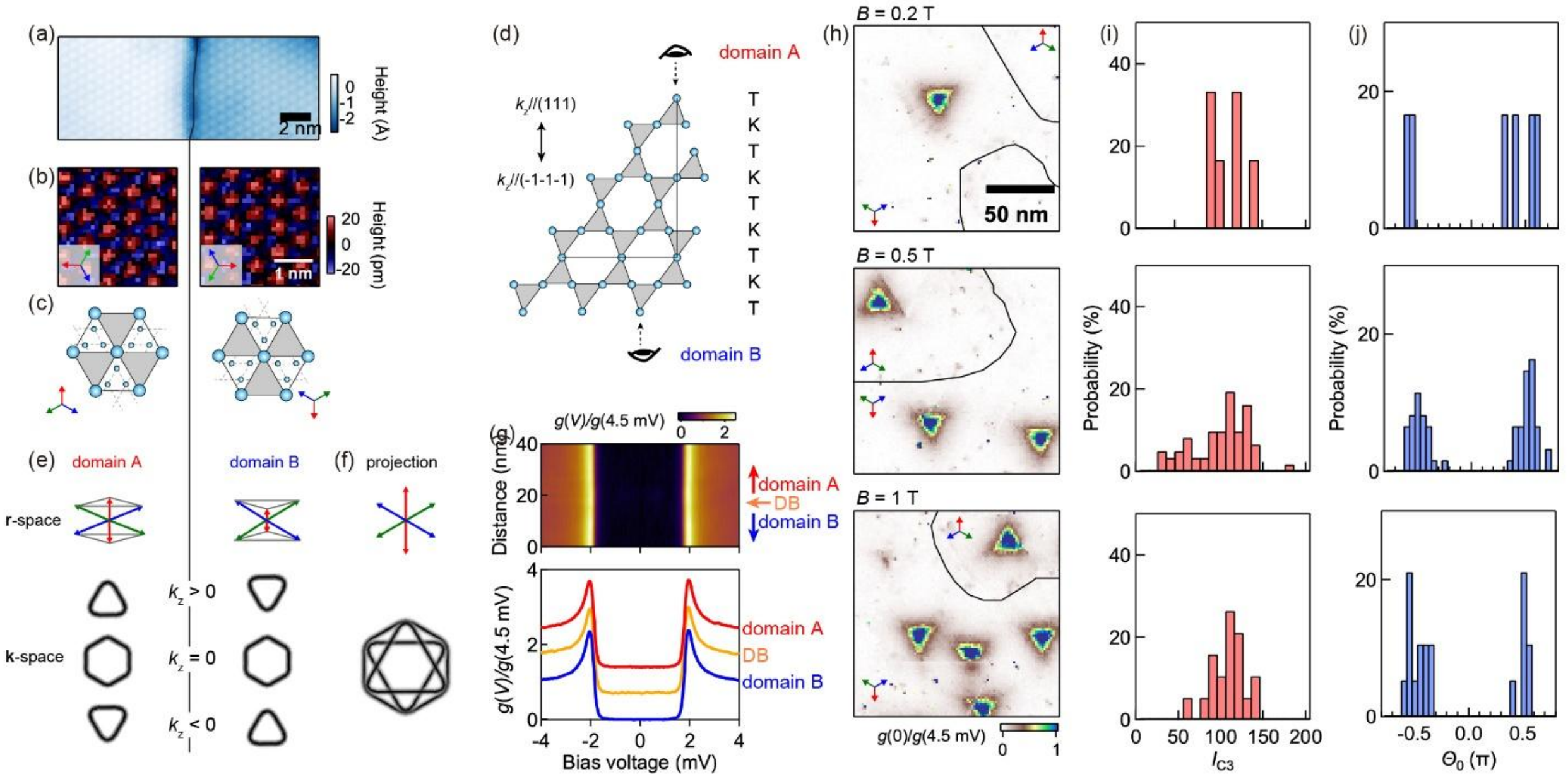

**FIG. 3. Triangular Abrikosov vortices in LTO.**

(a) STM topograph across a domain boundary (DB) with atomic resolution (1 V, 30 pA).

(b) The zoomed-in topographic image from two distinct domains. In addition to major red protrusions, which are visualized predominantly in Fig. 1(e), two types of local minima are visualized by blue protrusions (lower and shallower).

(c) Schematic Ti sublattice viewed along the (111) direction.

(d) Schematic illustration of the pyrochlore Ti lattice from the side view of the experimental geometry. The notation K and T represent the kagome and the triangular lattices of Ti, respectively.

(e,f) Schematic Fermi surfaces with varying kz for two structural domains (e) and their identical projection onto (111) plane (f).

(g) Spatial evolution of g(V)/g(4.5 mV) across a DB (g), and representative spectra taken in two distinct domains and on the DB at B = 0 T (h). Setup: +7 mV, 1 nA; lock-in: 0.2 mV, 961 Hz.

(h) ZBC maps taken under different magnetic fields of 0.2, 0.5, and 1 T. Setup: 5 mV, 500 pA; lock-in: 0.1 mV, 961 Hz.

(i,j) Histogram of $I_{C3}$ and $\theta_0$ for different magnetic fields, respectively.

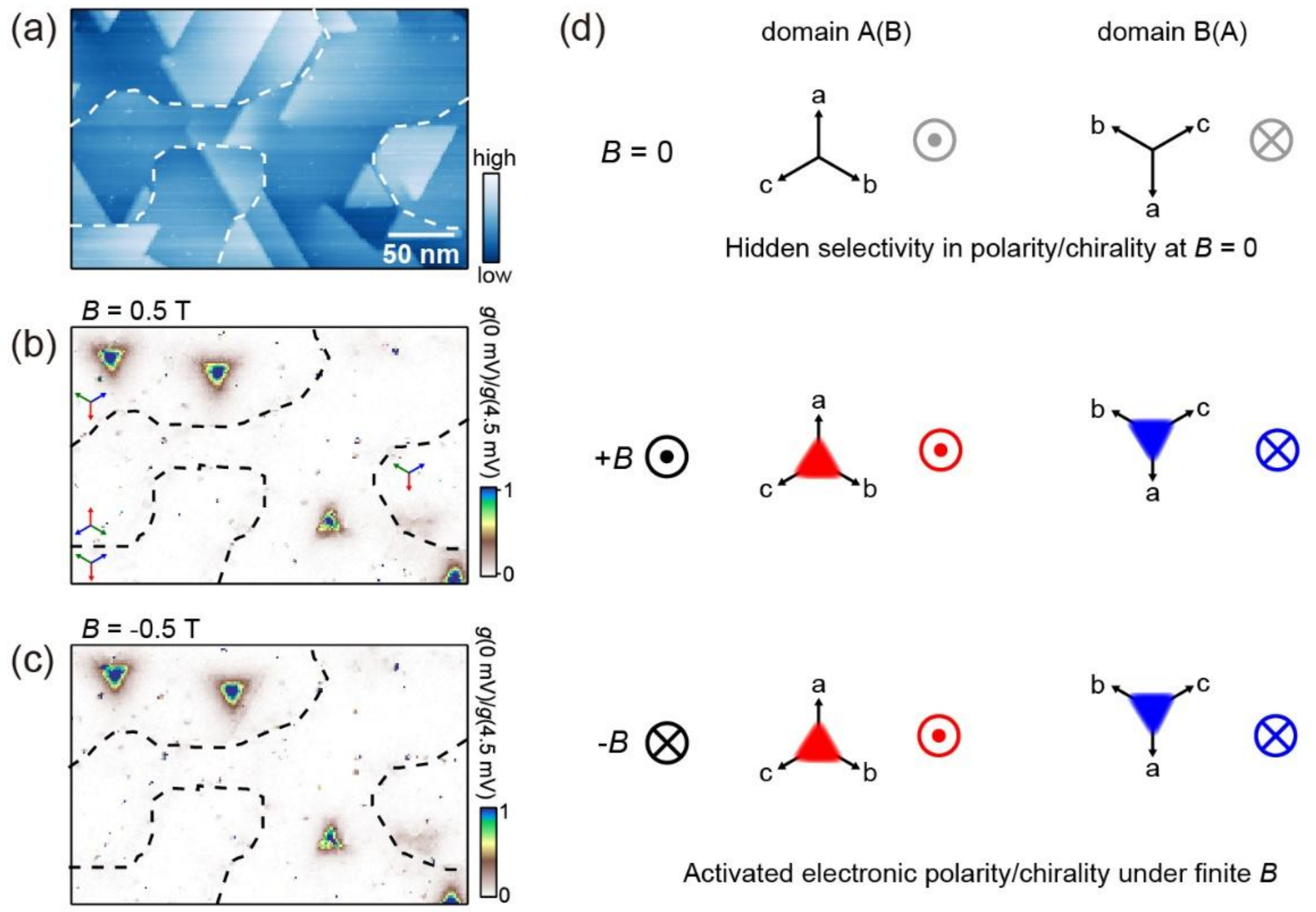

**Figure 4. B-direction independence of tr-AV orientation and hidden chirality in LTO.**
(a) STM image where conductance maps (b,c) are taken. Setup: 5 mV, 500 pA; lock-in: 0.1 mV, 961 Hz.
(b,c) Zero-bias conductance maps measured under magnetic fields of +0.5 T and −0.5 T, respectively.
(d) Schematic illustration of the hidden polarity/chirality selectivity of the triangular vortex state. At zero magnetic field, opposite vortex polarities associated with different domains are degenerate. Under finite magnetic field, the degeneracy is lifted, resulting in magnetic-field-selective stabilization of opposite triangular vortex configurations.